\documentclass[12pt]{iopart}
\usepackage{graphicx}
\usepackage{bm}

\begin{document}

\title[Evolving wormhole geometries within nonlinear electrodynamics]
{Evolving wormhole geometries within nonlinear electrodynamics}

\author{Aar\'{o}n V. B. Arellano\footnote[1]{vynzds@yahoo.com.mx}}
\address{Facultad de Ciencias,
Universidad Aut\'{o}noma del Estado de M\'{e}xico, \\
El Cerrillo, Piedras Blancas, C.P. 50200, Toluca, M\'{e}xico}
\address{and Centro de Astronomia
e Astrof\'{\i}sica da Universidade de Lisboa,\\
Campo Grande, Ed. C8 1749-016 Lisboa, Portugal}

\author{Francisco S. N. Lobo\footnote[2]{flobo@cosmo.fis.fc.ul.pt}}
\address{Centro de Astronomia
e Astrof\'{\i}sica da Universidade de Lisboa,\\
Campo Grande, Ed. C8 1749-016 Lisboa, Portugal}


\begin{abstract}

In this work, we explore the possibility of evolving $(2+1)$ and
$(3+1)-$dimensional wormhole spacetimes, conformally related to
the respective static geometries, within the context of nonlinear
electrodynamics.
For the $(3+1)-$dimensional spacetime, it is found that the
Einstein field equation imposes a contracting wormhole solution
and the obedience of the weak energy condition. Nevertheless, in
the presence of an electric field, the latter presents a
singularity at the throat, however, for a pure magnetic field the
solution is regular. For the $(2+1)-$dimensional case, it is also
found that the physical fields are singular at the throat. Thus,
taking into account the principle of finiteness, which states that
a satisfactory theory should avoid physical quantities becoming
infinite, one may rule out evolving $(3+1)-$dimensional wormhole
solutions, in the presence of an electric field, and the
$(2+1)-$dimensional case coupled to nonlinear electrodynamics.

\end{abstract}



\maketitle

\section{Introduction}

Nonlinear electrodynamics has recently found many applications in
several branches, namely, as effective theories at different
levels of string/M-theory \cite{Witten}, cosmological models
\cite{cosmo-models}, black holes \cite{blackholes,Bronnikov1}, and
in wormhole physics \cite{Bronnikov1,wormhole1,wormhole2}, amongst
others. Pioneering work on nonlinear electrodynamic theories may
be traced back to Born and Infeld \cite{BI}, where the latter
outlined a model to remedy the fact that the standard picture of a
point charged particle possesses an infinite self-energy. Thus,
the Born-Infeld model was founded on a principle of finiteness,
that a satisfactory theory should avoid physical quantities
becoming infinite. Furthermore, Pleba\'{n}ski extended the
examples of nonlinear electrodynamic Lagrangians \cite{Pleb}, and
demonstrated that the Born-Infeld theory satisfy physically
acceptable requirements.

In this context, in a recent paper, it was shown that $(2+1)$ and
$(3+1)-$dimensional static, spherically symmetric and stationary,
axisymmetric traversable wormholes cannot be supported by
nonlinear electrodynamics \cite{wormhole2}. This is mainly due to
the presence of an event horizon and that the null energy
condition is not violated. However, a particularly interesting
situation arouse in the analysis of the $(2+1)-$dimensional static
and spherically symmetric wormholes, namely, that in order to
construct these geometries, they must necessarily be supported by
physical fields that become singular at the throat. Thus, taking
into account the principle of finiteness, and imposing a
non-singular behaviour of the physical quantities, it was found
that the wormhole possesses an event horizon, rendering the
geometry non-traversable. We also point out that the non-existence
of $(3+1)-$dimensional static and spherically symmetric
traversable wormholes is consistent with previous results
\cite{Bronnikov1}.

Based on this analysis, it is also of interest to explore the
specific case of evolving wormhole geometries, in the context of
nonlinear electrodynamics. Time-dependent spherically symmetric
wormholes have been extensively analysed in the literature, and a
particularly interesting case of a dynamic wormhole immersed in an
inflationary background, was considered by Roman~\cite{Roman}. The
primary goal in the Roman analysis was to use inflation to enlarge
an initially small, possibly submicroscopic, wormhole, and test
whether one could evade the violation of the energy conditions in
the process. Further dynamic wormhole geometries were analysed,
considering specific cases \cite{Kar1,Kar2,Trobo}, and the
evolution of a wormhole model was considered in a FRW background
\cite{Kim-evolvWH}. In the latter model, it was found that the
total stress-energy tensor does not necessarily violate the energy
conditions as in the static Morris-Thorne case
\cite{Morris,Visser}. Different scenarios for the weak energy
condition violation were also explored, namely for a constant
redshift function \cite{Kar1,Kar2}, and an extension with a
specific choice of a non-zero redshift function was further
analysed \cite{Trobo}.

Thus, in this work, we shall be interested in exploring the
possibility that nonlinear electrodynamics may support
time-dependent traversable wormhole geometries. This is of
particular interest as the energy conditions are not necessarily
violated in specific regions, as noted above. Therefore,
analogously to Ref. \cite{wormhole2}, we shall consider $(2+1)$
and $(3+1)-$dimensional spacetimes and find particularly
interesting results. We find that for the $(3+1)-$dimensional case
in the presence of an electrical field, the latter becomes
singular at the throat, however, for a purely magnetic field, the
solution is regular at the throat, which is extremely promising,
and is in close relationship to the regular magnetic black holes
coupled to nonlinear electrodynamics found in Ref.
\cite{Bronnikov1}. For the $(2+1)-$dimensional case we find that
the physical fields become singular at the throat.

This paper is outlined in the following manner: In section
\ref{Sec:4dynWH} we analyze $(3+1)-$dimensional dynamic and
spherically symmetric wormholes coupled with nonlinear
electrodynamics, and in section \ref{Sec:3dynWH},
$(2+1)-$dimensional evolving wormhole geometries in the context of
nonlinear electrodynamics are studied. In section \ref{Conclusion}
we conclude. We shall use geometrised units, i.e., $G=c=1$,
throughout this work
.

\section{$(3+1)-$dimensional wormhole}\label{Sec:4dynWH}

\subsection{Action and spacetime metric}

The action of $(3+1)-$dimensional general relativity coupled to
nonlinear electrodynamics is given by
\begin{equation}
S=\int \sqrt{-g}\left[\frac{R}{16\pi}+L(F)\right]\,d^4x  \,,
\end{equation}
where $R$ is the Ricci scalar. $L(F)$ is a gauge-invariant
electromagnetic Lagrangian, depending on a single invariant $F$
given by $F=F^{\mu\nu}F_{\mu\nu}/4$ \cite{Pleb}, where
$F_{\mu\nu}$ is the electromagnetic tensor. Note that in
Einstein-Maxwell theory, the Lagrangian takes the form $L(F)=
-F/4\pi$, however, we shall consider more general choices for the
electromagnetic Lagrangians.

Varying the action with respect to the gravitational field
provides the Einstein field equations $G_{\mu\nu}=8\pi
T_{\mu\nu}$, with the stress-energy tensor given by
\begin{equation}
T_{\mu\nu}=g_{\mu\nu}\,L(F)-F_{\mu\alpha}F_{\nu}{}^{\alpha}\,L_{F}\,,
    \label{4dim-stress-energy}
\end{equation}
where $L_F=dL/dF$. The variation with respect to the
electromagnetic potential $A_\mu$, where
$F_{\mu\nu}=A_{\nu,\mu}-A_{\mu,\nu}$, yields the electromagnetic
field equations
\begin{equation}
\left(F^{\mu\nu}\,L_{F}\right)_{;\mu}=0 \,.
     \label{em-field}
\end{equation}

We shall consider that the spacetime metric representing a dynamic
spherically symmetric $(3+1)-$dimensional wormhole, which is
conformally related to the static wormhole geometry \cite{Morris},
takes the form
\begin{equation} \label{4dysme}
ds^2=\Omega^2(t)\left[-e ^{2\Phi(r)}dt^2+\frac{dr^2}{1-
b(r)/r}+r^2(d\theta ^2+\sin ^2{\theta}d\phi ^2)\right]  \,,
\end{equation}
where $\Phi$ and $b$ are functions of $r$, and $\Omega=\Omega(t)$
is the conformal factor, which is finite and positive definite
throughout the domain of $t$. $\Phi$ is the redshift function, and
$b(r)$ is denoted the form function \cite{Morris}. We shall also
assume that these functions satisfy all the conditions required
for a wormhole solution, namely, $\Phi(r)$ is finite everywhere in
order to avoid the presence of event horizons; $b(r)/r<1$, with
$b(r_0)=r_0$ at the throat; and the flaring out condition
$(b-b'r)/b^2 \geq 0$, with $b'(r_0)<1$ at the throat.

Now, taking into account metric (\ref{4dysme}), the
electromagnetic tensor, compatible with the symmetries of the
geometry, is given by
\begin{equation}
F_{\mu\nu}=E(x^\alpha)\;(\delta^t_\mu \delta^r_\nu-\delta^r_\mu
\delta^t_\nu)+B(x^\alpha)\;(\delta^\theta_\mu
\delta^\phi_\nu-\delta^\phi_\mu \delta^\theta_{\nu})
\label{4em-tensor}\,,
\end{equation}
where the non-zero components are the following:
$F_{tr}=-F_{rt}=E$, the electric field, and
$F_{\theta\phi}=-F_{\phi\theta}=B$, the magnetic field.
The invariant $F$, included for self-completeness, takes the
following form
\begin{equation}
F=-\frac{1}{2\Omega^4}\left[\left(1-\frac{b}{r}\right)e^{-2\Phi}\,E^2-\frac{B^2}{r^4
\sin^2\theta}\right] \,.  \label{invF}
\end{equation}

\subsection{Mathematics of embedding}

To analyze evolving wormhole spacetimes, $\Phi(r)$ and $b(r)$ are
chosen to provide a plausible wormhole geometry at $t=0$, which is
assumed to be the onset of the evolution. One may verify the
evolution of the wormhole considering the proper circumference
$C_0$ of the throat, $r_0$, given by
\begin{equation}
C_0=  \int_{0}^{2\pi} \Omega(t)\,r_0\,d\phi= \Omega(t)2\pi r_0 \,,
\end{equation}
and the radial proper length through the wormhole, between any two
points $A$ and $B$ at any $t={\rm const}$, provided by
\begin{equation}
l(t)=\pm \Omega(t) \int_{r_A}^{r_B}
\frac{dr}{(1-b/r)^{1/2}}=\Omega(t)\,l(t=0) \,,
\label{evolvproperdistance}
\end{equation}
which is simply the product of $\Omega(t)$ and the initial proper
circumference and the radial proper separation, respectively.

One may use the mathematics of embedding to verify that the
wormhole form of the metric is preserved with time. Here, we shall
closely follow the analysis outlined in Ref. \cite{Roman}. Due to
the spherically symmetric nature of the problem, one may, without
a significant loss of generality, consider an equatorial slice
$\theta=\pi/2$. Considering, in addition, a fixed moment of time,
$t={\rm const}$, the metric of the wormhole slice is given by
\begin{equation}\label{slice}
ds^2={\Omega^2(t)\,{dr^2}\over{1-b(r)/r}} + \Omega^2(t)\,
r^2\,d\phi^2\,.
\end{equation}
Now, to visualize this slice, this metric is embedded in a flat
three dimensional Euclidean space with metric
\begin{equation}
ds^2=d{\bar{z}}^2+d{\bar{r}}^2+{\bar{r}}^2\,{d\phi}^2\,.
\label{barredslice}
\end{equation}
Comparing the respective coefficients of ${d\phi}^2$, one verifies
the following relationships
\begin{eqnarray}
\bar{r}={\Omega(t)\,r}\big|_{t={\rm const}} \,,
       \qquad
{d\bar{r}}^2=\Omega^2(t)\,{dr}^2\big|_{t={\rm const}}  \,.
\label{coef:phi}
\end{eqnarray}
However, it is important to emphasize, in particular, when
considering derivatives, that equations (\ref{coef:phi}) do not
represent a coordinate transformation, but rather a rescaling of
the $r$ coordinate on each $t={\rm constant}$ slice \cite{Roman}.

Note that the wormhole form of the metric will be preserved if the
embedded metric, written in $\bar{z}$, $\bar{r}$ and $\phi$
coordinates, has the form
\begin{equation}\label{WHslice}
ds^2={{d{\bar{r}}^2}\over{1-{\bar{b}(\bar{r})/{\bar{r}}}}} +
                      {\bar{r}}^2{d\phi}^2\,,
\end{equation}
where $\bar{b}(\bar{r})$ has a minimum at some
$\bar{b}(\bar{r}_0)=\bar{r}_0$. Equation (\ref{slice}) can be
rewritten in the form of equation (\ref{WHslice}) by using
equations (\ref{coef:phi}) and the definition
$\bar{b}(\bar{r})=\Omega(t)\,b(r)$.

Using equations (\ref{barredslice}) and (\ref{WHslice}), one
deduces $\bar{z}(\bar{r})=\Omega(t)\,z(r)$. Note that the
relationship between the embedding space at an arbitrary time $t$
and the initial embedding space at $t=0$, using the above
expressions, is given by
\begin{equation}
ds^2=d{\bar{z}}^2+d{\bar{r}}^2+{\bar{r}}^2\,{d\phi}^2
        =\Omega^2(t)\,[dz^2+dr^2+r^2{d\phi}^2]\,.
\end{equation}
Relative to the $({\bar{z}},{\bar{r}},\phi)$ coordinate system the
wormhole will always remain the same size, as the scaling of the
embedding space compensates for the evolution of the wormhole.
However, the wormhole will change size relative to the initial
$t=0$ embedding space \cite{Roman}.

An important aspect of wormhole physics is the flaring out
condition. From the above analysis, it follows that
\begin{equation}
{{d\,^2{\bar{r}(\bar{z})}}\over{d{\bar{z}}^2}}
       =\frac{1}{\Omega(t)}\,{{b-b'r}\over{2b^2}}
       =\frac{1}{\Omega(t)}\,{{d\,^2r(z)}\over{dz^2}}>0\,,
       \label{barred:flareout}
\end{equation}
at or near the throat \cite{Morris}, so that the flaring out
condition for the evolving wormhole is given by
$d\,^2{\bar{r}(\bar{z})/d\bar{z}}^2>0$, in order to provide a
wormhole solution. Taking into account equation (\ref{coef:phi}),
the definition of $\bar{b}(\bar{r})$, and
$\bar{b}'(\bar{r})=d\bar{b}/d\bar{r}=b'(r)=db/dr$, one may rewrite
equation (\ref{barred:flareout}) relative to the barred
coordinates as $d\,^2{\bar{r}(\bar{z})}/d{\bar{z}}^2
=\bar{b}-\bar{b}'\bar{r}/(2{\bar{b}}^2)>0$, at or near the throat.
Thus, using barred coordinates, the flaring out condition has the
same form as for the static wormhole \cite{Morris}.

\subsection{Einstein field equations}

For convenience, the non-zero Einstein tensor components, in an
orthonormal reference frame, are given in \ref{A:4Einstein}, and
the stress-energy tensor components, using equation
(\ref{4dim-stress-energy}), in \ref{A:4SET}. Now, using equation
(\ref{4Totr}), i.e., $T_{\hat t\hat r}=0$, and the Einstein field
equation, $G_{\hat\mu\hat\nu}=8\pi T_{\hat\mu\hat\nu}$, we verify
from equation (\ref{4Gotr}) that $\Phi'=0$, considering the
non-trivial case $\dot\Omega \neq 0$. Without a significant loss
of generality, we choose $\Phi=0$, so that the nonzero components
of the Einstein tensor, equations (\ref{4Gott})-(\ref{4Gohhpp})
reduce to
\begin{eqnarray}
G_{\hat t\hat
t}&=&\frac{1}{\Omega^2}\left[\frac{b'}{r^2}+3\left(\frac{\dot{\Omega}}{\Omega}\right)^2\right]\,,
      \label{4rGott}  \\
G_{\hat r\hat r}&=&\frac{1}{\Omega^2}\left\{-\frac{b}{r^3}+
  \left[\left (\frac{\dot{\Omega}}{\Omega}\right
)^{2}-2\,\frac{\ddot {\Omega}}{\Omega} \right]\right\}
  \,,  \label{4rGorr}   \\
G_{\hat\theta\hat\theta}&=&G_{\hat\phi\hat\phi}=\frac{1}{\Omega^2}\left\{-\frac{b'r-b}{2r^3}
  +\left[\left (\frac{\dot{\Omega}}{\Omega}\right
)^{2}-2\,\frac{\ddot {\Omega}}{\Omega} \right]\right\}  \,,
             \label{4rGohhpp}
\end{eqnarray}
where the overdot denotes a derivative with respect to the time
coordinate, $t$, and the prime a derivative with respect to $r$.
Note that the metric for this particular case is identical to the
specific metric analyzed in Ref. \cite{Kar1}.

Finally, taking into account that $\Phi=0$, the nonzero components
of the stress energy tensor, from equations
(\ref{4Tott})-(\ref{4Tohhpp}), take the following form
\begin{eqnarray} \label{4rset}
T_{\hat t\hat t}&=&-L-\frac{(1-b/r)}{\Omega^4}E^2L_{F}\,,
    \label{4rTott}   \\
T_{\hat r\hat r}&=&L+\frac{(1-b/r)}{\Omega^4}E^2L_{F}\,,
    \label{4rTorr}   \\
T_{\hat\theta\hat\theta}&=&T_{\hat\phi\hat\phi}=L-\frac{1}{\Omega^4r^4\sin^2\theta}B^2L_{F}\,.
    \label{4rTohhpp}
\end{eqnarray}
To impose the finiteness of the stress-energy tensor components,
we shall also impose that $|(1-b/r)\,E^2\,L_F/\Omega^4|<\infty$
and $|B^2\,L_F/(\Omega^4r^4\sin^2\theta)|<\infty$, as
$r\rightarrow r_0$.

An interesting feature immediately stands out, namely, that
$T_{\hat t\hat t}=-T_{\hat r\hat r}$, so that using equations
(\ref{4rGott})-(\ref{4rGorr}) through the Einstein field equation,
we obtain the following relationship
\begin{equation}\label{4boeq}
\frac{b'r-b}{2r^3}=-\left[2\left(\frac{\dot\Omega}{\Omega}\right)^2
-\frac{\ddot\Omega}{\Omega}\right]
  \,.
\end{equation}
Now, equation (\ref{4boeq}) can be solved separating variables and
provides the following solutions,
\begin{equation}\label{4solb}
  b(r)=r\left[1-\alpha^2(r^2-r_0^2)\right]    \,,
\end{equation}
\begin{equation}\label{4solom}
  \Omega(t)=\frac{2\alpha}{C_1\,e^{\alpha t}-C_2\,e^{-\alpha t}}
  \,,
\end{equation}
where $\alpha$ is a constant, and $C_1$ and $C_2$ are constants of
integration. Note that the form function reduces to $b(r_0)=r_0$
at the throat, and $b'(r_0)=1-2\alpha^2 r_0^2<1$ is also verified
for $\alpha \neq 0$. Relatively to the conformal function, if
$C_1=C_2$, then $\Omega$ is singular at $t=0$.

Thus, if $\alpha>0$, then we need to impose $C_1e^{\alpha
t}>C_2e^{-\alpha t}$, and if $\alpha<0$ and $C_1e^{\alpha
t}<C_2e^{-\alpha t}$, otherwise the conformal factor becomes
singular somewhere along the domain of $t$. Now, $\Omega(t)
\rightarrow 0$ as $t\rightarrow \infty$, which reflects a
contracting wormhole solution. This analysis shows that one may,
in principle, obtain an evolving wormhole solution in the range of
the time coordinate.
Note that $\alpha$ has the dimensions of (distance)$^{-1}$, so
that it will be useful to define a dimensionless parameter
$\beta=\alpha r_0$, so that the form function is given by
\begin{equation}\label{4solb2}
b(r)=r\left\{1-\beta^2\left[\left(\frac{r}{r_0}\right)^2-1\right]\right\}\,.
\end{equation}
A fundamental condition to be a solution of a wormhole, is that
$b(r)>0$ is imposed \cite{wormhole-shell}. Thus, the range of $r$
is $r_0<r<a=r_0\sqrt{1+1/\beta^2}$, and the latter may be made
arbitrarily large by taking $\beta \rightarrow 0$. If $a \gg r_0$,
i.e., $\beta\simeq r_0/a\ll 1$, one may have an arbitrarily large
wormhole. Note, however, that one may, in principle, match this
solution to an exterior vacuum solution at a junction interface
$R$, within the range $r_0<r<a$, much in the spirit of Refs.
\cite{wormhole-shell,wormhole-shell2}.

It is also important to point out an interesting physical feature
of this evolving, and in particular, contracting geometry, namely,
the absence of the energy flux term, $T_{\hat t\hat r}=0$. One can
interpret this aspect considering that the wormhole material is at
rest in the rest frame of the wormhole geometry, i.e., an observer
at rest in this frame is at constant $r,\,\theta,\,\phi$. The
latter coordinate system coincides with the rest frame of the
wormhole material, which can be defined as the one in which an
observer co-moving with the material sees zero energy flux. This
analysis is similar to the one outlined in Ref. \cite{Roman}.

In conclusion to this section, note that the solution outlined
above possesses solely electromagnetic fields. One could also
consider a non-interacting anisotropic time-dependent distribution
of matter coupled to nonlinear electrodynamics. This may be
reflected by the following superposition of the stress-energy
tensor
\begin{equation}
T_{\mu\nu}=T_{\mu\nu}^{\rm fluid}+T_{\mu\nu}^{\rm NED} \,,
\end{equation}
where $T_{\mu\nu}^{\rm NED}$ is given by Eq.
(\ref{4dim-stress-energy}), and $T_{\mu\nu}^{\rm fluid}$ is the
anisotropic time-dependent stress-energy tensor associated with
the fluid. A first approach reveals formidable calculations due to
the time-dependence of the solution. Nevertheless, this is an
interesting case to explore, and in particular the analysis of the
energy conditions, which we leave for future work.

\subsection{Energy conditions}

\begin{figure}[h]
\centering
  \includegraphics[width=2.8in]{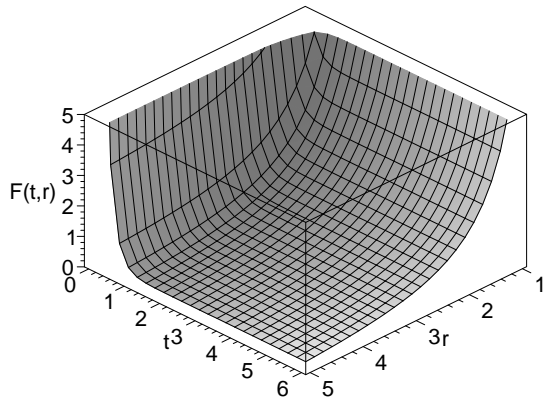}
  \caption{Plot of the inequality (\ref{ineq1}), which is given by
  the region above the surface. The latter is defined by
  $F(r,t)=[b'/r^2+3(\dot\Omega/\Omega)^2]$, thus
  satisfying inequality (\ref{ineq1}). See the text for details.}
  \label{fig:wec}
\end{figure}

We may further explore the energy conditions much in the spirit of
Ref. \cite{Kar1}, in particular, the weak energy condition (WEC),
which is defined as
$T_{\hat\mu\hat\nu}U^{\hat\mu}U^{\hat\nu}\geq0$ where
$U^{\hat\mu}$ is a timelike vector. The fact that the stress
energy tensor is diagonal will be helpful, so that we need only to
check the following three conditions
\begin{eqnarray}
T_{\hat t\hat t}\geq0 \,,   \qquad
T_{\hat t\hat t}+T_{\hat r\hat r}\geq0 \,,  \qquad
T_{\hat t\hat t}+T_{\hat\theta\hat\theta}\geq0  \,,
\end{eqnarray}
which using equations (\ref{4rGott})-(\ref{4rGohhpp}), provide the
following inequalities
\begin{eqnarray}
\frac{1}{\Omega^2}\left[\frac{b'}{r^2}+3\left(\frac{\dot\Omega}{\Omega}\right)^2\right]\geq0
\,,   \label{ineq1}
\\
\frac{1}{\Omega^2}\left\{\frac{b'r-b}{2r^3}+
\left[2\left(\frac{\dot\Omega}{\Omega}\right)^2-\frac{\ddot\Omega}{\Omega}\right]\right\}
\geq 0 \,,    \label{ineq2} \\
\frac{1}{\Omega^2}\left\{\frac{b'r+b}{2r^3}+
2\left[2\left(\frac{\dot\Omega}{\Omega}\right)^2-\frac{\ddot\Omega}{\Omega}\right]\right\}
\geq 0 \,.    \label{ineq3}
\end{eqnarray}
Note that inequality (\ref{ineq2}) reduces to the null energy
condition (NEC) for a null vector oriented along the radial
direction \cite{Kar1} (The NEC is defined as
$T_{\hat\mu\hat\nu}k^{\hat\mu}k^{\hat\nu}\geq0$ where
$k^{\hat\mu}$ is a null vector). In fact, an interesting feature
for the present $(3+1)-$dimensional evolving wormhole geometry
coupled to nonlinear electrodynamics, is that the NEC is zero, for
a null vector oriented along the radial direction. This may be
inferred from equation (\ref{4boeq}), i.e.,
$T_{\hat\mu\hat\nu}k^{\hat\mu}k^{\hat\nu}=0$, for arbitrary $t$
and $r$. For inequality (\ref{ineq3}), using equations
(\ref{4boeq}) and (\ref{4solb2}), we obtain $2r(1+\beta^2)\geq0$
which is always fulfilled. Finally, inequality (\ref{ineq1}),
which is graphically depicted in figure \ref{fig:wec}, is also
satisfied. We have defined
$F(r,t)=[b'/r^2+3(\dot\Omega/\Omega)^2]$, which represents the
surface plotted in figure \ref{fig:wec}. The inequality
(\ref{ineq1}) is represented as the region above the surface, and
is manifestly positive.

We emphasize that for a static wormhole geometry the null energy
condition, i.e., condition (\ref{ineq2}), is necessarily violated,
consequently implying the violation of the WEC. However, this is
not the case for dynamic wormhole spacetimes, as already pointed
out in Ref. \cite{Kar1}. In the context of nonlinear
electrodynamics, it was shown that $(3+1)-$dimensional static and
spherically symmetric traversable wormholes cannot exist
\cite{Bronnikov1,wormhole2}, as the NEC is not violated, so that
the flaring out condition is not verified. However, for the case
of the evolving wormhole geometry analysed in this work, we have
verified that the WEC is satisfied, as shown in the analysis
above.

In the context of the energy conditions, a general class of higher
dimensional wormhole geometries were constructed in an interesting
paper \cite{DeBen}, in which the four non-compact dimensions are
static, and the possibility of time-dependent compact extra
dimensions was explored. An interior wormhole solution was matched
to an exterior vacuum solution, using the Synge junction
conditions. The results of the analysis showed that: firstly, for
the static case, where the gravitational field does not evolve in
the full space, it is possible to respect the WEC at the throat,
provided the extra dimensions place a restriction on the radial
size of the wormhole throat; secondly, for the quasi-static case,
where only the compact dimensions are time-dependent, the WEC
cannot be satisfied at the throat. The latter result differs from
the analysis within the context of non-linear electrodynamics
explored in this work, where the time-dependent wormhole
geometries satisfy the WEC. This is due to the fact that the
higher dimensional solutions analysed in Ref. \cite{DeBen}, the
matter field does not possess an infinite spatial extent, due to
the matter-vacuum junction boundary. As the models are now matched
to the exterior vacuum, the time dependence of the extra
dimensions is fixed by the vacuum and cannot de chosen
arbitrarily, consequently resulting in the violation of the WEC
(see Ref. \cite{DeBen} for further details).

\subsection{Electromagnetic field equations}

The electromagnetic field equation, equation (\ref{em-field}), for
calculational convenience, may be rewritten as
\begin{equation}
(F^{\mu\nu}L_F)_{,\mu}=-F^{\mu\nu}\,\Gamma^{\alpha}_{\;\mu\alpha}\,L_F
\,, \label{4emfeq}
\end{equation}
where $\Gamma^{\alpha}_{\;\mu\nu}$ are the Christoffel symbols of
the second type.
Setting $\nu=t$ and $\nu=r$, respectively, equation (\ref{4emfeq})
can be solved, yielding the following solution for $EL_{F}$
\begin{equation}\label{4ELF}
  EL_{F}=\frac{C_{E}}{r^2(1-b/r)^{1/2}}  \,,
\end{equation}
where $C_{E}$ is a constant or a function of $\theta$ only. From
this last relation we verify that $EL_{F}$ is independent of the
$t$ coordinate, and it is singular at the throat.
Analogously, setting $\nu=\phi$, equation (\ref{4emfeq}) can be
solved, to provide the following relationship for $BL_{F}$
\begin{equation}\label{4BLF}
  BL_{F}=C_{B}(t,r)\,\Omega^4\,r^4\,\sin\theta  \,,
\end{equation}
where $C_{B}$ is a constant of integration.

Another relationship, fundamental to our analysis, is the
following
\begin{equation}\label{hodge}
\left(^*F^{\mu\nu}\right)_{;\mu}=0 \,,
\end{equation}
which can be deduced from the Bianchi identities, where $^*$
denotes the hodge dual \cite{Bronnikov1}. Now, from equation
(\ref{hodge}), we obtain $\dot B=0$, $B'=0$ and $E_{,\theta}=0$.
Then, these conditions, together with equations (\ref{4ELF}) and
(\ref{4BLF}) give us that $F_{tr}=-F_{tr}=E(t,r)$,
$F_{\theta\phi}=-F_{\phi\theta}=B(\theta)$, $L_{F}=L_{F}(t,r)$.
Thus, one may take $C_E=q_{\rm e}={\rm const}$, and note that the
magnetic field is given by
\begin{equation}\label{4NB}
  B(\theta)=q_{\rm m}\,\sin\theta  \,,
\end{equation}
where $q_{\rm e}$ and $q_{\rm m}$ are constants related to the
electric and magnetic charge, respectively.

Furthermore, from equations (\ref{4rGott}), (\ref{4rGohhpp}),
(\ref{4rTott}) and (\ref{4rTohhpp}), we obtain
\begin{equation}\label{4rTtmh}
  \frac{\Omega^2}{8\pi}\left(\frac{b'r-3b}{2r^3}\right)=\left(1-\frac{b}{r}\right)\,E^2\,L_{F}+
  \frac{q_{\rm m}^2}{r^4}L_{F}\,.
\end{equation}
Considering a non-zero electric field, $E\neq0$, we can use
equations (\ref{4ELF}) and (\ref{4rTtmh}) to obtain
\begin{equation}\label{4NE}
  E(t,r)=\frac{(b'r-3b)\Omega^2r\pm\sqrt{(b'r-3b)^2\Omega^4r^2
  -(32\pi q_{\rm e} q_{\rm m})^2}}{32\pi q_{\rm e}
  r^2(1-b/r)^{1/2}}\,.
\end{equation}
From this solution we point out two observations: (i) we require
that $(b'r-3b)^2\Omega^2r^2>(32\pi q_{\rm e}q_{\rm m})^2$ so we
have a limiting (interval) inequality; (ii) and that $E$ is
inversely proportional to $(1-b/r)^{-1/2}$, showing that the $E$
field is singular at the throat, which is in contrast to the
principle of finiteness. Finally, and for completeness, we have
\begin{equation}\label{4NLF}
  L_{F}=\frac{32q_{\rm e}^2\pi}{(b'r-3b)\Omega^2r\pm\sqrt{(b'r-3b)^2\Omega^4r^2
  -(32\pi q_{\rm e} q_{\rm m})^2}}\,.
\end{equation}

\subsubsection{$B=0$.}

In particular, consider the case of $B=0$, so that using equations
(\ref{4rGott})-(\ref{4rGorr}), (\ref{4rTott})-(\ref{4rTorr}) and
(\ref{4boeq}), we find
\begin{equation}\label{4E2LF}
  E^2L_{F}=\frac{\Omega^2}{16\pi}\frac{b'r-3b}{r^3(1-b/r)} \,,
\end{equation}
which together with equation (\ref{4ELF}) provides
\begin{equation}\label{4E}
  E=\frac{\Omega^2}{16\pi q_{\rm e}}\frac{b'r-3b}{r(1-b/r)^{1/2}} \,,
\end{equation}
and
\begin{equation}\label{4LF}
  L_{F}=\frac{16\pi q_{\rm e}^2}{\Omega^2r(b'r-3b)}  \,.
\end{equation}
Note that even if $B=0$ the expression for $E$ is singular at the
throat.

In the analysis outlined above, namely, in the presence of an
electric field, we verify a problematic issue, namely, that the
latter presents a singularity at the throat. This is an extremely
troublesome aspect of the geometry, and we emphasize that this is
in contradiction to the model construction of nonlinear
electrodynamics, founded on a principle of finiteness, that a
satisfactory theory should avoid physical quantities becoming
infinite \cite{BI}. Thus, one should impose that these physical
quantities be non-singular, and in doing so, we may rule out
$(3+1)-$dimensional dynamical spherically symmetric wormhole
solutions, in the presence of electric fields, within the context
of nonlinear electrodynamics.

\subsubsection{$E=0$.}

An interesting case arises setting $E=0$. Using equation
(\ref{4rTtmh}), we obtain
\begin{equation}\label{4NBLF}
  L_{F}=\frac{1}{16\pi q_{\rm m}^2}\Omega^2r(b'r-3b)\,,
\end{equation}
and taking into account equations (\ref{4rGott}) and
(\ref{4rTott}), the Lagrangian is given by
\begin{equation}\label{4NBL}
  L=-\frac{1}{8\pi\Omega^2}\left[\frac{b'}{r^2}+3\left(\frac{\dot{\Omega}}{\Omega}\right)^2\right]\,.
\end{equation}
These equations, together with $B=q_{\rm m}\sin\theta$, $E=0$,
$F=q_{\rm m}^2/(2\Omega^4r^4)$ and solutions (\ref{4solb}) and
(\ref{4solom}) give a wormhole solution without problems at the
throat, with finite fields. This result is in close relationship
to the regular magnetic black holes coupled to nonlinear
electrodynamic found in Ref. \cite{Bronnikov1}.

\section{$(2+1)-$dimensional wormhole}\label{Sec:3dynWH}

\subsection{Action and spacetime metric}

We shall also be interested in $(2+1)-$dimensional general
relativity coupled to nonlinear electrodynamics. The respective
action is given by
\begin{equation}
S=\int \sqrt{-g}\left[\frac{R}{16\pi}+L(F)\right]\,d^3x  \,,
\end{equation}
where $L(F)$ is a gauge-invariant electromagnetic Lagrangian,
which we shall leave unspecified at this stage, depending on a
single invariant $F$ given by $F=F^{\mu\nu}F_{\mu\nu}/4$. The
factor $1/16\pi$ is maintained in the action to keep the
parallelism with $(3+1)-$dimensional theory. The Maxwell
Lagrangian is recovered in the weak field limit, i.e.,
$L(F)\rightarrow -F/4\pi$. Analogously to the $(3+1)-$dimensional
case, by varying the action with respect to the gravitational
field, one obtains the Einstein field equations $G_{\mu\nu}=8\pi
T_{\mu\nu}$, where the stress-energy tensor is given by
\begin{equation}
T_{\mu\nu}=g_{\mu\nu}\,L(F)-F_{\mu\alpha}F_{\nu}{}^{\alpha}\,L_{F}\,.
    \label{stress-energy}
\end{equation}
Varying the action with respect to the electromagnetic potential,
one obtains the electromagnetic field equation,
$(F^{\mu\nu}\,L_F)_{;\mu}=0$.

Consider the time-dependent spherically symmetric
$(2+1)-$dimensional wormhole spacetime, and which is given by the
following metric
\begin{equation}\label{dyswh}
ds^2=\Omega^2(t)\left[-e^{2\Phi}dt^2+\frac{dr^2}{1-b/r}+r^2d\phi^2\right]
  \,.
\end{equation}

Taking into account the symmetries of the geometry, we shall
consider the following electromagnetic tensor
\begin{equation}
F_{\mu\nu}=E(x^\alpha)(\delta^t_\mu \delta^r_\nu-\delta^r_\mu
\delta^t_\nu)+B(x^\alpha)(\delta^\phi_\mu
\delta^r_\nu-\delta^r_\mu \delta^\phi_{\nu})
\label{em-tensor}\,,
\end{equation}
where the nonzero components are given by
$F_{tr}=-F_{rt}=E(x^\alpha)$ and $F_{\phi
r}=-F_{r\phi}=B(x^\alpha)$. We shall include the expression for
the invariant $F$, for self-completeness, which is given by
\begin{equation}
F=-\frac{1}{2}\,\frac{(1-b/r)}{\Omega^2}\,\left(e^{-2\Phi}
\,E^2-\frac{1}{r^2}B^2\right) \,.
\end{equation}

\subsection{Field equations}

The non-zero Einstein tensor components are given in
\ref{B:3Einstein}, and the respective stress-energy tensor
components in \ref{B:3SET}. Analogously for the
$(3+1)-$dimensional case, note that $T_{\hat t\hat r}=0$, and
using the Einstein field equation, $G_{\hat\mu\hat\nu}=8\pi
T_{\hat\mu\hat\nu}$, we verify from equation (\ref{Gotr}) that
$\Phi'=0$, considering the non-trivial case $\dot\Omega \neq 0$.
Once again, one may consider $\Phi=0$, without a loss of
generality, so that the nonzero components of the Einstein tensor,
equations (\ref{Gott})-(\ref{Gopp}), may be rewritten as
\begin{eqnarray}
G_{\hat t\hat t}&=&\frac{1}{\Omega^2}\left[\frac{b'r-b}{2r^3}
+\left(\frac{\dot{\Omega}}{\Omega}\right)^2\right]\,,
        \label{rGott}  \\
G_{\hat r\hat
r}&=&G_{\hat\phi\hat\phi}=\frac{1}{\Omega^2}\left[\left(\frac{\dot\Omega}
{\Omega}\right)^2-\frac{\ddot{\Omega}}{\Omega}\right]\,,
        \label{rGorr}
\end{eqnarray}
and the respective nonzero components of the stress energy tensor,
equations (\ref{Tott})-(\ref{Topp}), take the form
\begin{equation}\label{rTott}
T_{\hat t\hat t}=-L-\frac{(1-b/r)}{\Omega^4}E^2L_{F} \,,
\end{equation}
\begin{equation}\label{rTorr}
   T_{\hat r\hat
  r}=L+\frac{(1-b/r)}{\Omega^4}E^2L_{F}-\frac{(1-b/r)}{r^2\Omega^4}B^2L_{F} \,,
\end{equation}
\begin{equation}\label{rTopp}
  T_{\hat\phi\hat\phi}=L-\frac{(1-b/r)}{r^2\Omega^4}B^2L_{F}  \,.
\end{equation}
Furthermore, from equations (\ref{rGorr}) and
(\ref{rTorr})-(\ref{rTopp}), using the Einstein field equation we
verify that $E=0$, ignoring the trivial case $L_{F}=0$. Note that
from $T_{\hat t \hat\phi}$, equation (\ref{Totp}) one has $E\cdot
B=0$, which is consistent with $E=0$, as found above. Thus, the
stress energy components reduce to
\begin{eqnarray}
T_{\hat t\hat t}&=&-L    \,,  \label{TottII}   \\
T_{\hat r\hat r}&=&T_{\hat\phi\hat\phi}=
L-\frac{(1-b/r)}{r^2\Omega^4}B^2L_{F} \,, \label{ToppII}
\end{eqnarray}
and the Lagrangian is given by
\begin{eqnarray}
L=-\frac{1}{8\pi \Omega^2}\left[\frac{b'r-b}{2r^3}
+\left(\frac{\dot{\Omega}}{\Omega}\right)^2\right]\,.
\end{eqnarray}
We can now calculate the WEC, by taking into account equations
(\ref{rGott})-(\ref{rGorr}), to obtain the following relationships
\begin{eqnarray}\label{dysNECg}
\frac{1}{\Omega^2}\left[\frac{b'r-b}{2r^3}
+\left(\frac{\dot{\Omega}}{\Omega}\right)^2\right] \geq 0\,,
          \\
\frac{1}{\Omega^2}\left\{\frac{b'r-b}{2r^3}+
\left[2\left(\frac{\dot\Omega}
{\Omega}\right)^2-\frac{\ddot{\Omega}}{\Omega}\right]\right\} \geq
0 \,.
\end{eqnarray}

From the electromagnetic field equations, we obtain $F^{\mu
t}{}_{;\mu}=0$ and $F^{\mu r}{}_{;\mu}=0$ and
\begin{equation}\label{emfep}
\frac{(1-b/r)}{r^2\Omega^4}\,B\,\left[\log(L_{F})\right]_{,r}
=F^{\mu\phi}{}_{;\mu}
\,.
\end{equation}
Equation (\ref{emfep}) can be solved to provide
\begin{equation}\label{BLFII}
BL_{F}=\frac{C_t r}{(1-b/r)^{1/2}} \,,
\end{equation}
where $C_t$ is a constant or a function of $t$ only. Furthermore,
from equations (\ref{TottII})-(\ref{ToppII}) we obtain
\begin{equation}\label{B2LFII}
  B^2L_{F}=-\frac{\Omega^2r^2}{8\pi(1-b/r)}\left\{\frac{b'r-b}{2r^3}+
  \left[2\left(\frac{\dot\Omega}
{\Omega}\right)^2-\frac{\ddot{\Omega}}{\Omega}\right]\right\} \,.
\end{equation}
Thus, using equations (\ref{BLFII}) and (\ref{B2LFII}), we find
\begin{equation}\label{BII}
  B(t,r)=-\frac{1}{8\pi C_t}\frac{\Omega^2r}{(1-b/r)^{1/2}}\left\{\frac{b'r-b}{2r^3}+
  \left[2\left(\frac{\dot\Omega}
{\Omega}\right)^2-\frac{\ddot{\Omega}}{\Omega}\right]\right\}  \,,
\end{equation}
and
\begin{equation}\label{LFII}
L_{F}(t,r)=-\frac{8\pi C_t^2}{\Omega^2}
\left\{\frac{b'r-b}{2r^3}+\left[2\left(\frac{\dot\Omega}
{\Omega}\right)^2-\frac{\ddot{\Omega}}{\Omega}\right]\right\}^{-1}
\,.
\end{equation}
From equation (\ref{BII}), one verifies that the field $B$ is
singular at the throat. Analogously to the $(3+1)-$ dimensional
case, this is an extremely troublesome aspect of the geometry, as
in order to construct a traversable wormhole, singularities appear
in the physical fields, which is in contradiction to the model
construction of nonlinear electrodynamics, founded on a principle
of finiteness \cite{BI}. Thus, one should impose that these
physical quantities be non-singular, and in doing so, we verify
that we cannot afford a wormhole type solution.

\section{Conclusion}\label{Conclusion}

In a recent paper, it was shown that $(2+1)$ and
$(3+1)-$dimensional static, spherically symmetric and stationary,
axisymmetric traversable wormholes cannot be supported by
nonlinear electrodynamics. In this work, we explored the
possibility of evolving time-dependent wormhole geometries coupled
to nonlinear electrodynamics. For the $(3+1)-$dimensional
spacetimes, it was found that the Einstein field equation imposes
a contracting wormhole solution and that the weak energy condition
be satisfied. It was also found that in the presence of an
electric field, a problematic issue was verified, namely, that the
latter become singular at the throat. However, regular solutions
of traversable wormholes in the presence of a pure magnetic field
were found.
Time-dependent spherically symmetric $(2+1)-$dimensional wormhole
spacetimes were also analyzed, and it was found that the Einstein
field equation imposes that the electric field be zero. For this
case, it was found that in order to construct wormhole geometries,
these must necessarily be supported by physical fields that become
singular at the throat. Thus, taking into account that the model
construction of nonlinear electrodynamics, founded on the
principle of finiteness, that a satisfactory theory should avoid
physical quantities becoming infinite, one may rule out evolving
$(3+1)-$dimensional electric wormhole solutions, and the
$(2+1)-$dimensional case coupled to nonlinear electrodynamics.

It is also relevant to emphasize that the solutions obtained in
this work and in Ref. \cite{wormhole2}, can be obtained using an
alternative form of nonlinear electrodynamics, denoted the $P$
framework \cite{Bronnikov1}. The latter is obtained from the
original form, the $F$ framework, by a Legendre transformation.
The duality between the $F$ and $P$ frameworks connects solutions
of different theories, but we emphasize that it is a dual
description of the same physical system. Therefore, we have not
made use of the $P$ formalism throughout this work, as we have
only been interested in exploring the possible existence of
evolving wormhole solutions coupled to nonlinear electrodynamics.
Another point worth noting is that we have only considered that
the gauge-invariant electromagnetic Lagrangian $L(F)$ be dependent
on a single invariant $F$. As stressed in Ref. \cite{wormhole2},
it would also be worthwhile to include another electromagnetic
field invariant $G\sim \,^*F^{\mu\nu}\,F_{\mu\nu}$, which would
possibly add an interesting analysis to the solutions found in
this work.





\appendix
\section{$(3+1)-$dimensional evolving wormhole geometry}
\label{A:4dim-wormhole}

\subsection{Einstein tensor}\label{A:4Einstein}

The non-zero components of the Einstein tensor, given in an
orthonormal reference frame, for the metric (\ref{4dysme}), are
the following
\begin{eqnarray}
\fl G_{\hat t\hat t}=\frac{1}{\Omega^2}\,\left[ 3
e^{-2\Phi}\,\left(\frac{\dot {\Omega}}{\Omega}\right)^2 +\frac{b
'}{r^2} \right] \,, \label{4Gott}
       \\
\fl G_{\hat r\hat r}=\frac{1}{\Omega ^{2}}\;\left\{
e^{-2\Phi(r)}\,\left[{\left ({{\dot {\Omega}} \over
{\Omega}}\right )^{2}}-2\,\frac{\ddot {\Omega}}{\Omega} \right]
- \left[{b \over
{r^{3}}}-2\,\frac{\Phi'}{r}\,\left(1-\frac{b}{r}\right) \right]
\right\}             \,,    \label{4Gorr}
       \\
\fl G_{\hat t\hat
r}=2\,\frac{\dot{\Omega}}{\Omega^3}\;e^{-\Phi}\,\Phi'
\left(1-\frac{b}{r} \right)^{1/2}        \,, \label{4Gotr}
       \\
\fl G_{\hat\theta\hat\theta}=G_{\hat\phi\hat\phi}= \frac{1}{\Omega
^{2}}\;\Bigg\{
e^{-2\Phi(r)}\,\left[\left (\frac{\dot{\Omega}}{\Omega}\right
)^{2}-2\,\frac{\ddot {\Omega}}{\Omega} \right]
+ \left(1-\frac{b}{r}\right) \times
         \nonumber    \\
\times\left[\Phi ''+ (\Phi')^2 - \frac{b'r-b}{2r(r-b)}\Phi'-
\frac{b'r-b}{2r^2(r-b)}+\frac{\Phi'}{r} \right] \Bigg\} \,,
\label{4Gohhpp}
\end{eqnarray}
where the overdot denotes a derivative with respect to the time
coordinate, $t$, and the prime a derivative with respect to $r$.

\subsection{Stress-energy tensor}\label{A:4SET}

The components of the stress energy tensor, equation
(\ref{4dim-stress-energy}), in the orthonormal frame, take the
following form
\begin{eqnarray} \label{4set}
T_{\hat t\hat
t}&=&-L-\frac{e^{-2\Phi}(1-b/r)}{\Omega^4}E^2L_{F}\,,
    \label{4Tott}   \\
T_{\hat r\hat r}&=&L+\frac{e^{-2\Phi}(1-b/r)}{\Omega^4}E^2L_{F}\,,
    \label{4Torr}   \\
T_{\hat\theta\hat\theta}&=&T_{\hat\phi\hat\phi}=L-\frac{1}{\Omega^4r^4\sin^2\theta}B^2L_{F}\,,
    \label{4Tohhpp}   \\
T_{\hat t\hat i}&=&T_{\hat i \hat j}=0 \quad ({\rm with} \;\; i
\neq j)\,.
    \label{4Totr}
\end{eqnarray}


\section{$(2+1)-$dimensional evolving wormhole geometry}
\label{B:3dim-wormhole}

\subsection{Einstein tensor}\label{B:3Einstein}

Using the orthonormal reference frame we have that the nonzero
components of the Einstein tensor, for the metric (\ref{dyswh}),
are
\begin{eqnarray}
\fl G_{\hat t\hat
t}&=&\frac{1}{\Omega^2}\left[\frac{b'r-b}{2r^3}+e^{-2\Phi}
\left(\frac{\dot{\Omega}}{\Omega}\right)^2\right]\,,
      \label{Gott}  \\
\fl G_{\hat r\hat
r}&=&\frac{1}{\Omega^2}\left\{\left(1-\frac{b}{r}\right)\frac{\Phi'}{r}+
  e^{-2\Phi}\left[\left(\frac{\dot\Omega}{\Omega}\right)^2
  -\frac{\ddot{\Omega}}{\Omega}\right]\right\}
  \,,  \label{Gorr}   \\
\fl
G_{\hat\phi\hat\phi}&=&\frac{1}{\Omega^2}\Bigg\{\left(1-\frac{b}{r}\right)
\left[\Phi''+(\Phi')^2- \frac{b'r-b}{2r(r-b)}\Phi' \right]
+e^{-2\Phi}\left[\left(\frac{\dot\Omega}{\Omega}\right)^2
-\frac{\ddot{\Omega}}{\Omega}\right]\Bigg\} \,,
             \label{Gopp}   \\
\fl G_{\hat t\hat
r}&=&\frac{1}{\Omega^2}\left[\left(1-\frac{b}{r}\right)^{1/2}\Phi'\;
e^{-\Phi}\left(\frac{\dot\Omega}{\Omega}\right)\right] \,.
      \label{Gotr}
\end{eqnarray}

\subsection{Stress-energy tensor}\label{B:3SET}

The components of the stress energy tensor, equation
(\ref{stress-energy}), in the orthonormal frame, take the
following form
\begin{eqnarray}
T_{\hat t\hat t}&=&-L-\frac{e^{-2\Phi}(1-b/r)}{\Omega^4}E^2L_{F}
\,,
       \label{Tott}    \\
T_{\hat r\hat
r}&=&L+\frac{e^{-2\Phi}(1-b/r)}{\Omega^4}E^2L_{F}-\frac{(1-b/r)}{r^2\Omega^4}B^2L_{F}
  \,,    \label{Torr}    \\
T_{\hat\phi\hat\phi}&=&L-\frac{(1-b/r)}{r^2\Omega^4}B^2L_{F}
                 \,,    \label{Topp}    \\
T_{\hat t\hat r}&=&0   \,,  \label{Totr}   \\
T_{\hat t\hat\phi}&=&-\frac{e^{-\Phi}(1-b/r)}{\Omega^2
r}\,\,E\,B\,L_F   \,, \label{Totp} \\
T_{\hat r\hat\phi}&=&0   \,.
\end{eqnarray}



\end{document}